\documentclass[a4paper,12pt]{article}
\usepackage{amsfonts,amsthm,amsmath,amssymb,graphicx,hyperref,youngtab,lmodern}
\newcommand{\tr}{\operatorname{tr}}

\def\Dim{\textrm{Dim}}

\def\mC{ \mathbb{C}}

\newcommand\stiny[1]{\fontsize{2.5}{4}\selectfont #1 }
\newcommand\mytiny[1]{\fontsize{4.5}{4}\selectfont #1 }
\newcommand\ybox{ \mytiny{\yng(1)}}
\newcommand\subybox{\stiny{ \yng(1)}}

\newcommand{\cA}{\mathcal A}

\newcommand{\cO}{\mathcal O}

\newcommand{\be}{\begin{equation}}
\newcommand{\bea}{\begin{eqnarray}}
\newcommand{\ee}{\end{equation}}
\newcommand{\eea}{\end{eqnarray}}

\def\Dim{{ \rm{Dim}} }

\begin{document}

\makeatletter
\@addtoreset{equation}{section}
\makeatother
\renewcommand{\theequation}{\thesection.\arabic{equation}}

\rightline{QMUL-PH-18-16}
\vspace{0.8truecm}

\vspace{15pt}

%%%%%%%%%%%%%%%%%

{\LARGE{
\centerline{ \bf  Quantum Information Processing }
\centerline{\bf   and Composite Quantum Fields.  }
}}

\vskip.5cm

\thispagestyle{empty} \centerline{
   {\large \bf   Sanjaye Ramgoolam${}^{a,b,}$\footnote{ {\tt s.ramgoolam@qmul.ac.uk}},
    Michal Sedl\'ak${}^{c,d, }$\footnote{\tt fyzimsed@savba.sk}
       }}

\vspace{.4cm}
\centerline{{\it ${}^a$ Centre for Research in String Theory, Department of Physics},}
\centerline{{ \it Queen Mary University of London},} \centerline{{\it    Mile End Road, London E1 4NS, UK}}

\vspace{.2cm}
\centerline{{\it ${}^b$ National Institute for Theoretical Physics,}}
\centerline{{\it School of Physics and Mandelstam Institute for Theoretical Physics,}}
\centerline{{\it University of the Witwatersrand, Wits, 2050, South Africa } }

\vspace{.2cm}
\centerline{ {\it ${}^c$ RCQI, Institute of Physics, Slovak Academy of Sciences,}}
\centerline{ {\it  Dubravska cesta 9, 84511 Bratislava, Slovakia}}

\centerline{ {\it ${}^d$ Faculty of Informatics,~Masaryk University,}}
\centerline{ {\it Botanick\'a 68a,~60200 Brno,~Czech Republic}}

\vspace{1.4truecm}

%%%%%%%%%%%%%%%%%
\thispagestyle{empty}

\centerline{\bf ABSTRACT}

\vskip.4cm

Some beautiful identities involving hook contents of Young diagrams
have been found in the field of quantum information processing, along with
a combinatorial proof. We here give a representation theoretic proof of these identities
and a number of generalizations. Our proof is based on trace identities for elements belonging to
a class of permutation centralizer algebras. These algebras have been found to underlie the combinatorics
of composite gauge invariant operators in quantum field theory, with applications in the AdS/CFT correspondence.
Based on these algebras, we discuss some analogies between  quantum information processing tasks and  the combinatorics of composite quantum fields and argue that this can be fruitful interface between quantum information and quantum field theory, with implications for AdS/CFT.

\setcounter{page}{0}
\setcounter{tocdepth}{2}

\newpage

\tableofcontents

\section{ Identities from Quantum Information Theory }

Some beautiful symmetric group identities have been found in the
subject of quantum information processing \cite{BSZ}.
A combinatoric proof has been given \cite{BS}.

The key identity is
\bea\label{EQ1}
d_r n ( n+1) = \sum_{ R \vdash (n+1)  } d_R ~ g ( r ,~ \ybox ~ , R ) ~  ( c_{ \stiny{ \yng(1)} } ( R , r ) )^2 \; ,
\eea

where $r$ is a Young diagram with $n$ boxes. $R$ is a Young diagram with $ n+1$ boxes,
$d_r$ is the dimension of the  irreducible representation (irrep)  of the symmetric group  $S_n$
associated with the Young diagram $r$, $d_R$ is the dimension of the $S_{ n+1} $ irrep
associated with $R$. Finally, $ g ( r, \ybox , R )$ is the Littlewood-Richardson coefficient
coupling the $ V^U_r \otimes V^U_{\subybox} $ with $ V^U_R$, where $V^U_{ r}, V^U_{ \subybox} , V^U_{ R } $ are the
$U(N)$ (or $GL(N)$) irreps associated with the respective Young diagrams and $c_{\subybox}(R,r)$ is
the content (difference of the column and row number) of the additional box in Young diagram $R$ which is not contained in $r$.

\section{ Representation theoretic proof }

Consider the tensor product of $U(N)$ irreps.
\bea
V_{ r}^{ U }  \otimes V^{ U}_{  \subybox }
\eea
It is an irredicible representation of $U(N) \times U(N)$.
Under the diagonal subgroup, it is reducible. The decomposition is
given by
\bea
V_r^{U} \otimes V_{ \subybox }^{U} = \bigoplus_{ R \vdash n +1 } g ( r , \ybox  , R   )   V_{ R }^{U}
\eea

It follows that
\bea\label{UNrbox}
N ~ \Dim_N ( r )  = \sum_{ R \vdash n +1 }  g ( r , \ybox , R  ) \Dim_N ~ R
\eea
We also have a representation of $ S_{n } \times S_1 $ associated
with $ ( r , \ybox )$. $S_{ n } \times S_1 $ is a subgroup of $S_{ n+1}$.
We can induce a representation of $S_{ n+1}$ from this representation of
$ S_n \times S_1$. This is a reducible representation of $S_{n+1}$.
The multiplicity of an irrep $R$ in this reducible rep is $ g ( r , \ybox , R ) $ (see for example \cite{FulHar}).

This means that
\bea\label{induced}
 ( n+1)  d_r  = \sum_{ R \vdash (n+1)  } g ( r , \ybox , R ) d_R
\eea
On the LHS we used the fact that the dimension of the rep of $S_{ n+1}$ induced
from the irrep $ V^{S_n}_r \otimes V^{S_1}_{\subybox} $ of $ S_n \times S_1$, is
\bea
{ | S_{ n+1} | \over  | S_n \times S_1 |} \Dim ( V^{S_n}_r \otimes V^{S_1}_{ \subybox} )
 = { ( n+1)! \over n! } d_r  = (n+1) d_r
\eea
On the RHS we use the decomposition in terms of irreps of $S_{n+1}$.

Useful relation between $ \Dim_N r $ and $d_r$ is
\bea\label{reldD}
\Dim_N r  = { d_r  f_r \over n! },
\eea
where $f_r$ is the product over the boxes of the Young diagram of $ ( N - c_{ \subybox } ) $ and
$c_{ \subybox} = i - j $ for a box at row $i$ and column $j$.
\bea
f_r = \prod_{ \ybox  ~ \in ~ r } ( N - c_{ \subybox} ) = \prod_{ i , j  }  ( N + j - i )
\eea
Similarly,
\bea
\Dim_N R = { d_R  f_R \over ( n+1) ! }.
\eea
Thus the ratio reads
\bea
{ \Dim_N R  \over \Dim_N r } = { d_R \over d_r ( n+1 )  }  ( N - c_{ \subybox} ( R , r ) ),
\eea
where $c_{ \subybox} ( R , r )$ is the content of the box by which $R$ and $r$ differ. We can also write
$c_{ \subybox} ( R , r )= \sum_{ \subybox \in R } c_{ \subybox} ( R ) - \sum_{ \subybox \in r } c_{ \subybox} ( r )$.
Using the above equation in (\ref{UNrbox}) we have
\bea
N =  \sum_{ R \vdash ( n+1) }  g ( r , \ybox , R ) ~~ { d_R \over d_r ( n+1 )  }  ( N - c_{ \subybox} ( R , r ) )
\eea
%Using this alongside
Comparing this with the induction equation (\ref{induced}), we find
\bea\label{vanishing}
 \sum_{ R \vdash ( n+1) }  g ( r , \ybox , R ) c_{ \subybox} ( R , r ) d_R  = 0.
\eea

\subsection{ Trace of a sum of permutations in $V_{ N}^{ \otimes ( n+1)}$   }

In the group algebra $ \mC ( S_{ n+m} ) $, an interesting sub-algebra
is formed by the subspace which is invariant under conjugation
by elements of $S_n \times S_m$. This is an example of what was called a  permutation centralizer algebra (PCA) in \cite{PCA,Kimura}, and which has many applications in the study of gauge invariant
operators with relevance to AdS/CFT.  It is
denoted $ \cA  ( n , m )$.  For  $m=1$ we have $ \cA ( n , 1 )$.
The element $ \sum_{ a =1}^{ n } ( a , n+1) $,
 which we will use here,  is an element of $ \cA ( n , 1 )$.
It is very interesting that PCAs are also finding a  use in quantum information processing (QIP).
We discuss this further in Section \ref{sec:Discussion}.

The tensor product
$V_r \otimes V_{ \subybox} $ is a subspace of $ V_{ N}^{ \otimes ( n+1)} $, where $V_N$ is the fundamental
of $U(N)$.
\bea
W =  V_{N}^{ \otimes n +1 } &&  = V_N^{ \otimes n } \otimes V_N \cr
 &&  =\bigoplus_{ r \vdash n } ( V_r^{ U }  \otimes V_r^{ S_n }  )  \otimes  ( V_{ \subybox}^{U}
  \otimes V_{ \subybox}^{S_1} )
\eea
Consider the projector $P_r$ in the group algebra of $ S_n$, denoted $\mC(S_n)$
\bea
P_{ r} = { d_r  \over n!  } \sum_{ \sigma \in S_n } \chi_r ( \sigma ) \sigma
\eea
We choose an embedding $ \mC ( S_n ) \rightarrow  \mC ( S_{ n+1} ) $, where
$ S_n $ acts on $ \{ 1 , 2, \cdots , n \}$ and $ S_{n+1} $ acts on $\{ 1, \cdots , n , n+1 \}$,
and
construct an element in $ \mC ( S_{ n+1} ) $
\bea
(  { P_{ r} \over d_r }  \otimes 1  )  \sum_{ a =1}^n ( a , n+1)
\eea
We then consider  the trace in $W$
\bea
\tr_W (  { P_{ r} \over d_r }  \otimes 1  )  \sum_{ a =1}^n ( a , n+1)
\eea

This is a sum of permutations in $ S_n \subset S_{ n+1} $. Doing the multiplication of
$P_r$ with $ \sum_{ a =1}^n ( a , n+1) $ and taking the trace, we get
\bea
 n ~ \chi_r \left ( { 1 \over n!  } \sum_{ \sigma  } N^{ C_{ \sigma } } \sigma \right ) = n ~ \Dim_N r
\eea
See Appendix eq. \ref{trsumprod} for the derivation.

Let us consider another way to compute the same trace.
We observe that
\bea
 \sum_{ a =1}^n ( a , n+1) =  T_2^{ (S_{ n+1} ) }  - T_2^{ (S_n)},
\eea
where $T_2^{(S_n)} $ is the sum of all permutations in $S_n$ which consist of
 a single swop.
Using Eqs. (\ref{centralchar}) and (\ref{LRchar}) we get
\bea
\label{traceoft2}
&& tr_{ W} ( { P_r \over d_r }  \otimes 1 ) \left ( T_2^{ (S_{ n+1} ) }  - T_2^{ (S_n)}  \right )  =
\sum_{ R \vdash (n+1)  } tr_W P_{ R }  ( P_r \otimes 1 )
\left ( T_2^{ (S_{ n+1} ) }  - T_2^{ (S_n)} \right ) \cr
&& =\sum_{ R \vdash (n+1)  }
\left (  { \chi_R ( T_2^{ (S_{n+1} )} ) \over d_R }  - { \chi_r ( T_2^{ (S_{n} )} ) \over d_r } \right )
tr_W P_{ R }  ( { P_r  \over d_r } \otimes 1 ) \cr
&& = \sum_{  R \vdash (n+1)   } ( - c_{ \subybox} ( R , r )  ) g ( r , \ybox , R ) \Dim_N R
\eea
We made use of the fact
\bea
tr_W P_{ R }  ( { P_r  \over d_r } \otimes 1 )
= g ( r , \ybox , R ) \Dim_N R
\eea
This is a special case $ k =1, n_1 = n , n_2 =1$ of an identity \ref{KeyId}  derived in Appendix \ref{sec:App}.

Now we have
\bea
 n \Dim_N r = \sum_{  R \vdash (n+1)   } ( - c_{ \subybox} ( r , R )  ) g ( r , \ybox , R )  \Dim_N R.
\eea
Dividing by $ \Dim_N r $ on both sides we obtain
\bea
&& n = \sum_{  R \vdash (n+1)   } ( - c_{ \subybox} ( r , R ) ) g ( r , \ybox , R ) { d_R ( N - c_{\subybox}   ( r , R )  )  \over d_r ( n+1) }  \cr
&& = \sum_{  R \vdash (n+1)   }  - N c_{ \subybox} ( r , R ) g ( r , \ybox , R )d_R
+ \sum_{  R \vdash (n+1)   }   (  c_{ \subybox} ( r , R ) )^2  g ( r , \ybox , R ){ d_R \over (n+1) d_r }.
\eea
The first term in the above equation is zero due to Eq. (\ref{vanishing}).
Thus, we get
\bea\label{desiredId}
d_r n ( n+1) = \sum_{  R \vdash  (n+1)   }   (  c_{ \subybox} ( r , R ) )^2  g ( r , \ybox , R ) d_R,
\eea
which is the desired identity.

\section{ A generalization with $ S_m \times S_n$ }

Consider $ S_{ m } \times S_n \rightarrow S_{ m+n} $ and the vector space
%The tensor space
$ W = V_N^{ \otimes m + n  } $.

For Young diagrams $r, s $ with $m,n$ boxes respectively, consider
\bea
V_{ r}^U \otimes V_s^U = \bigoplus_{ R \vdash m +n  } g ( r , s , R ) V_R^U
\eea
which gives the decomposition into irreducible representations of
the $U(N)$ which acts diagonally on $ V_r^U \otimes V_s^U$.
First consider the dimension on both sides of the equation
\bea
\Dim_N r \Dim_N s = \sum_{ R \vdash (m+n)} g( r , s , R )  \Dim_N R ,
\eea
%This means, using (\ref{reldD}),
% \bea
%{ f_r f_s d_r d_s \over m! n! } = \sum_{ R \vdash (m+n)} g( r , s , R ){  f_R d_R \over ( m+n)! }
%\eea
%Rewriting
which is using Eq. (\ref{reldD}) rewritten as
\bea
{ ( m+n)! \over m! n! } d_r d_s = \sum_{ R \vdash (m+n)} g( r , s , R ) d_R  { f_R \over f_r f_s }
\eea
On the other hand, using induction of $V^{S_m}_r \otimes V^{S_n}_s $ from $ S_m \times S_n $ to $ S_{ m+n}$ we obtain
\bea
{ ( m+n)! \over m! n! } d_r d_s = \sum_{ R  } g ( r , s , R ) d_R
\eea
The expression
$ { f_R \over f_r f_s } $ has a large $N$ expansion $ 1 + \cO ( 1/N) $.
Comparing the two equations, we conclude that all the $1/N$ corrections
in
\bea
{ f_R \over f_r f_s }  =
{ \prod_{ \subybox \in R } ( 1 - {c_{ \subybox} ( R ) \over N } ) \over \prod_{ \subybox \in r } ( 1 - {c_{ \subybox} ( r ) \over N } )\prod_{ \subybox \in s } ( 1 - {c_{ \subybox} ( s ) \over N } )}
\eea
lead to identities. For example,
\bea\label{expandIds}
&& 0 = \sum_{ R \vdash (m+n)} g( r , s , R ) d_R  \left ( - \sum_{ \subybox \in R } c_{ \subybox} ( R )
+ \sum_{ \subybox \in r } c_{ \subybox} ( r )  + \sum_{ \subybox \in s } c_{ \subybox} ( s )   \right ) \cr
&& 0 = \sum_{ R \vdash (m+n)} g( r , s , R ) d_R
\bigl ( \sum_{ i_1 <  i_2 \in R    } c_{i_1}  ( R ) c_{ i_2 } ( R )  - \sum_{ i \in R } \sum_{ k \in r } c_i ( R ) c_k (r)
- \sum_{ i \in R } \sum_{ l \in s } c_i ( R ) c_l (s) \cr
&& + \sum_{ k \in  r } \sum_{ l \in s } c_k ( r) c_l ( s ) + \sum_{ k \in r } ( c_k (r) )^2
+ \sum_{ l \in s  } (c_{l} (s ))^2
+ \sum_{ k_1 <  k_2 \in r } c_{ k_1}  (r) c_{k_2  } ( r )
+ \sum_{ l_1  < l_2 \in s } c_{l_1}  (s) c_{l_2  } ( s )    \bigr ) \cr
&&
\eea
There will also be higher order equations: at each order in $1/N$  the equation involves Littlewood-Richardson coefficients and Young diagram contents, hence just  data pertaining to the symmetric groups.
All the equations arise from the large $N$ expansion of
\bea
0  = \sum_{ R \vdash (m+n)} g( r , s , R ) d_R
\left (  { f_R \over f_r f_s } - 1 \right ).
\eea
Now consider the trace
\bea
tr_W (  ( { P_r \over d_r}  \otimes { P_s \over d_s}  ) ( T_2^{ ( S_{ m+n} )} - T_2^{ ( S_m )} - T_2^{ (S_n ) } ) ).
\eea
We observe that
\bea
( T_2^{ ( S_{ m+n} )} - T_2^{ ( S_m )} - T_2^{ (S_n ) } )
= \sum_{ a =1}^{ m } \sum_{ b = m+1}^{ m+n} ( a,  b )
\eea
It follows that
\bea\label{trres1}
&& tr_W (  ( { P_r \over d_r}  \otimes { P_s \over d_s} ) ( T_2^{ ( S_{ m+n} )} - T_2^{ ( S_m )} - T_2^{ (S_n ) } ) ) \cr
&& = { 1  \over m! n! } \sum_{ \sigma_1 \in S_m } \sum_{ \sigma_2 \in S_n } \chi_r ( \sigma_1 ) \chi_s( \sigma_2 )
tr_W \left ( ( \sigma_1 \otimes \sigma_2 ) \sum_{ a =1}^{ m } \sum_{ b = m+1}^{ m+n} ( a,  b )    \right )
\cr
&& = { mn   \over m! n! } \sum_{ \sigma_1 \in S_m } \sum_{ \sigma_2 \in S_n } \chi_r ( \sigma_1 ) \chi_s( \sigma_2 )  N^{ C_{ \sigma_1} + C_{ \sigma_2} -1 } \cr
&& = N^{ -1} m n~  \Dim_N r ~  \Dim_N s
\eea
On the other hand, using (\ref{centralchar}), (\ref{contents})  and (\ref{KeyId}),  we derive
\bea\label{trres2}
&& tr_W (  ( { P_r \over d_r}  \otimes { P_s \over d_s}  ) ( T_2^{ ( S_{ m+n} )} - T_2^{ ( S_m )} - T_2^{ (S_n ) } ) ) \cr
&& = \sum_{ R \vdash (m+n)  } g ( r , s, R ) \Dim_N R \left (- \sum_{ \subybox \in R } c_{ \subybox} ( R ) + \sum_{ \subybox \in r } c_{ \subybox} ( r ) + \sum_{ \subybox \in s}  c_{ \subybox} ( s ) \right )
\eea
Comparing (\ref{trres1}) and (\ref{trres2}), we have
\bea\label{genId}
{  ( m+n)!  \over (m-1)! (n-1) ! }  = N  \sum_{ R \vdash (m+n)  } g ( r , s, R ) { d_R \over d_r d_s} { f_R \over f_r f_s}  \left ( - \sum_{ \subybox \in R }  c_{ \subybox} ( R ) + \sum_{ \subybox \in r }  c_{ \subybox} ( r ) + \sum_{ \subybox \in s }  c_{ \subybox} ( s ) \right )
\eea

Consider the large $N$ expansion.
\bea
{ f_R \over f_r f_s } = 1 + { 1 \over N }  \left ( - \sum_{ \subybox \in R }  c_{ \subybox} ( R ) + \sum_{ \subybox \in r }  c_{ \subybox} ( r ) + \sum_{ \subybox \in s }  c_{ \subybox} ( s ) \right )
+ \cO ( { 1 \over N^2 } )
\eea
Fraction ${ f_R \over f_r f_s }$ is the only term in the summand of the RHS in (\ref{genId}) which contains $N$ dependence.
Considering the order $N$ term of the RHS, we get zero using the first identity in (\ref{expandIds}).
Considering the constant term, we get
\bea
{  ( m+n)!  \over (m-1)! (n-1) ! } = \sum_{ R \vdash (m+n)  } g ( r , s, R ) { d_R \over d_r d_s}
 \left ( - \sum_{ \subybox \in R }  c_{ \subybox} ( R ) + \sum_{ \subybox \in r }  c_{ \subybox} ( r ) + \sum_{ \subybox \in s }  c_{ \subybox} ( s ) \right )^2 \; .
\eea
Equivalently,
\bea\label{newId}
{  ( m+n)!  \over (m-1)! (n-1) ! } d_r d_s
= \sum_{ R \vdash (m+n)  } g ( r , s, R ) d_R \left ( - \sum_{ \subybox \in R }  c_{ \subybox} ( R ) + \sum_{ \subybox \in r }  c_{ \subybox} ( r ) + \sum_{ \subybox \in s }  c_{ \subybox} ( s ) \right )^2 \; .
\eea
Now it is easy to see that this is a generalization of (\ref{desiredId}).

\section{Multi-partite generalization }\label{multipartite}

Consider %$ S_{ n_1 } \times S_{ n_2 } \ldots \times S_{ n_k }  \rightarrow S_{n_1 + n_2 + \ldots + n_k } $.
$ S_{ n_1 } \times \ldots \times S_{ n_k }  \rightarrow S_{n_1 +\ldots + n_k } $ and tensor space $ W = V_N^{ \otimes n_1 + \ldots + n_k } $.

For Young diagrams $r_1 , r_2 , \cdots , r_k $ with $n_1 , n_2 , \cdots , n_k$ boxes, we have
representations $V_{r_1}^U , V_{r_2}^U , \cdots, V_{ r_k}^U$ of $U(N)$. Considering the decomposition
of the tensor product under the diagonal action of $U(N)$, we have
\bea\label{gmulti}
V_{ r_1}^U \otimes V_{ r_2}^U \otimes \cdots \otimes V_{ r_k}^U
= \bigoplus_{ R \vdash n_1 + \cdots + n_k  } g ( r_1 , r_2 , \cdots , r_k ; R ) V_R^U
\eea
The multiplicities $g (r_1 , \cdots , r_k ; R  )$ can be expressed in terms of
Littlewood-Richardson coefficients. For example
\bea
g( r_1 , r_2 ,r_3 ; R ) = \sum_{ S \vdash n_1 + n_2  } g ( r_1 , r_2 ; S ) g  ( S , r_3 ;  R )
\eea

By considering the dimensions on the two sides of (\ref{gmulti}), we have
\bea
\Dim_N r_{1} \ldots \Dim_N r_{k} = \sum_{ R \vdash (n_1 + \ldots + n_k)} g( r_1 ,\ldots,r_k ; R )  \Dim_N R
\eea
%Using (\ref{reldD}) the abovre equation can be rewritten as
%\bea
%{ f_{r_1}\ldots f_{r_k} d_{r_1} \ldots d_{r_k} \over n_1! \ldots n_k! } = \sum_{ R \vdash (n_1+\ldots+n_k)} g( r_1 , \ldots %, r_k , R ){  f_R d_R \over (n_1+\ldots+n_k)! }
%\eea
%Rewriting
Using (\ref{reldD}) the above equation can be rewritten as
\bea
\label{n1nkeqcomp1}
{(n_1+\ldots+n_k)!  \over n_1! \ldots n_k! }d_{r_1} \ldots d_{r_k}  = \sum_{ R \vdash (n_1+\ldots+n_k)} g( r_1 , \ldots , r_k ; R )d_R {  f_R  \over  f_{r_1}\ldots f_{r_k} }
\eea

Now we switch to considering the induction of representations of symmetric groups associated with the above Young diagrams.  Using induction  of $V^{S_{n_1}}_{ r_1 } \otimes \ldots \otimes V^{S_{n_k}}_{r_k} $ from $ S_{ n_1 } \times \ldots \times S_{ n_k }$ to $S_{n_1 +\ldots + n_k }$ we have
\bea
\label{n1nkeqcomp2}
{(n_1+\ldots+n_k)!  \over n_1! \ldots n_k! }d_{r_1} \ldots d_{r_k}  = \sum_{ R %\vdash (n_1+\ldots+n_k)
} g( r_1 , \ldots , r_k ; R ) d_R
\eea
Comparing equations (\ref{n1nkeqcomp1}), (\ref{n1nkeqcomp2}) we have
\bea
\label{n1nkeqcomp3}
0 = \sum_{ R \vdash (n_1+\ldots+n_k)} g( r_1 , \ldots , r_k ; R ) d_R (1-{  f_R  \over  f_{r_1}\ldots f_{r_k} })
\eea
Again we consider a large $N$ expansion %$ 1 + \cO ( 1/N) $
of
\bea
{  f_R  \over  f_{r_1}\ldots f_{r_k} }  =
{ \prod_{ \subybox \in R } ( 1 - {c_{ \subybox} ( R ) \over N } ) \over \prod_{ \subybox \in r_1 } ( 1 - {c_{ \subybox} ( r_1 ) \over N } )\ldots \prod_{ \subybox \in r_k } ( 1 - {c_{ \subybox} ( r_k ) \over N } )}
\eea
%$ {  f_R  \over  f_{r_1}\ldots f_{r_k} }$ has
Validity of Eq. (\ref{n1nkeqcomp3}) for all $N \geq \sum_{j=1}^k n_j$ leads to identities for every power of $1/N$.
For example, the $1/N$ terms give
\bea\label{expandIdsn}
 0 = \sum_{ R \vdash (n_1+ \ldots +n_k)} g( r_1,\ldots,r_k , R ) d_R  \left ( - \sum_{ \subybox \in R } c_{ \subybox} ( R )
+ \sum_{i=1}^k \sum_{ \subybox \in r_i } c_{ \subybox} ( r_i )     \right )
\eea
%Another way to say it
%\bea
%0 = \sum_{ R \vdash (n_1+\ldots+n_k)} g( r_1 , \ldots , r_k , R )d_R \left ( {  f_R  \over  f_{r_1}\ldots f_{r_k} } - 1 \right )
%\eea
%Now consider
It is useful to be explicit about the embedding of $ S_{ n_1} \times \cdots S_{ n_k}$ in $ S_{ n_1 + \cdots n_k}$. Let $S_{ n_1} $ be the group of permutations of
$ [n_1] = \{ 1, 2, \cdots , n_1 \}$. Let $ S_{ n_2} $ be the group of permutations of
$ [ n_2] = \{ n_1 +1 , \cdots , n_1 + n_2 \}$. And $ S_{ n_i} $ for $ 1 \le i \le k $ be the
group of permutations of
\bea
[ n_i ]  = \{n_1 + n_2 + \cdots n_{ i-1 }  +1 , \cdots, n_1 + n_2 + \cdots + n_i \}
\eea
We also let $ S_{ n_1 + n_2 + \cdots n_1 + n_2 + \cdots n_k }$ be the group of
permutations of $ \{ 1, 2, \cdots , n_1 + n_2 + \cdots + n_k \}$.
Let us evaluate the trace
\bea
tr_W \left(  ( { P_{r_1} \over d_{r_1}}  \otimes \ldots \otimes { P_{r_k} \over d_{r_k}}  ) ( T_2^{ ( S_{n_1 + \ldots + n_k } )} - \sum_{i=1}^k T_2^{ ( S_{n_i} )}) \right)
\eea
in two ways.
We observe that
\bea
 T_2^{ ( S_{n_1 + \ldots + n_k} )} - \sum_{i=1}^k T_2^{ ( S_{n_i} )}
= \sum_{i=1}^{k-1} \; \sum_{j=i+1}^{k} \; \sum_{ a \in [ n_i ] }  \; \sum_{ b \in [ n_j ] } ( a,  b ) ,
\eea
Direct calculation (analogous to Eq.(\ref{trres2})) gives
\bea\label{trres1n}
&& tr_W \left(  ( { P_{r_1} \over d_{r_1}}  \otimes \ldots \otimes { P_{n_k} \over d_{n_k}}  ) ( T_2^{ ( S_{n_1 + \ldots + n_k } )} - \sum_{i=1}^k T_2^{ ( S_{n_i} )}) \right)
= { 1  \over n_1!\ldots n_k! } \sum_{ \sigma_1 \in S_{n_1} }\ldots \sum_{ \sigma_k \in S_{n_k} } \cr
&&
 \times \; \chi_{r_1} ( \sigma_1 )\ldots \chi_{r_k}( \sigma_k )\; tr_W \left ( ( \sigma_1 \otimes \ldots \otimes \sigma_k ) \sum_{i=1}^{k-1} \; \sum_{j=i+1}^{k} \; \sum_{ a \in [ n_i  ] }  \; \sum_{ b \in [ n_j ] }  ( a,  b )  \right )
\cr
&& = { \sum_{i=1}^{k-1} \; \sum_{j=i+1}^{k} n_{i} n_{j}   \over n_1!\ldots n_k! } \sum_{ \sigma_1 \in S_{n_1} }\ldots \sum_{ \sigma_k \in S_{n_k} } \chi_{r_1} ( \sigma_1 )\ldots \chi_{r_k}( \sigma_k ) \; N^{ C_{ \sigma_1} +\ldots+ C_{ \sigma_k} -1 } \cr
&& = { \sum_{i=1}^{k-1} \; \sum_{j=i+1}^{k} n_{i} n_{j} \over N } \;  \Dim_N r_1 \ldots \Dim_N r_k
\eea

On the other hand, using (\ref{centralchar}), (\ref{contents}) and (\ref{KeyId}), we have
\bea\label{trres2n}
&& tr_W \left(  ( { P_{r_1} \over d_{r_1}}  \otimes \ldots \otimes { P_{n_k} \over d_{n_k}}  ) ( T_2^{ ( S_{n_1 + \ldots + n_k } )} - \sum_{i=1}^k T_2^{ ( S_{n_i} )}) \right) \cr
&& =
\sum_{ R \vdash (n_1+ \ldots +n_k)} g( r_1,\ldots,r_k , R ) \Dim_N R  \left ( - \sum_{ \subybox \in R } c_{ \subybox} ( R )
+ \sum_{i=1}^k \sum_{ \subybox \in r_i } c_{ \subybox} ( r_i )     \right )
\eea

Comparing (\ref{trres1n}) and (\ref{trres2n}), we have
\bea\label{genIdn}
&& (\sum_{i=1}^{k-1} \; \sum_{j=i+1}^{k} n_{i} n_{j}) {(n_1 + \ldots + n_k)!   \over n_1!\ldots n_k! }  = \\
&& N  \sum_{ R \vdash (n_1+ \ldots +n_k)} g( r_1,\ldots,r_k , R ) {d_R \over d_{r_1}\ldots d_{r_k}} {f_R \over f_{r_1}\ldots f_{r_k}}  \left ( - \sum_{ \subybox \in R } c_{ \subybox} ( R )
+ \sum_{i=1}^k \sum_{ \subybox \in r_i } c_{ \subybox} ( r_i )     \right ) \nonumber
\eea

Consider the large $N$ expansion.
\bea
{f_R \over f_{r_1}\ldots f_{r_k}} = 1 + { 1 \over N } \left ( - \sum_{ \subybox \in R } c_{ \subybox} ( R )
+ \sum_{i=1}^k \sum_{ \subybox \in r_i } c_{ \subybox} ( r_i )     \right )
+ \cO ( { 1 \over N^2 } )
\eea
In Eq. (\ref{genIdn}) the only term in the summand of the RHS, which contains $N$ dependence is the fraction ${f_R \over f_{r_1}\ldots f_{r_k}}$.
Considering the order $N$ term of the RHS, we get zero using the first identity in (\ref{expandIdsn}).
Considering the constant term, we get
\bea
&& (\sum_{i=1}^{k-1} \; \sum_{j=i+1}^{k} n_{i} n_{j}) {(n_1 + \ldots + n_k)!   \over n_1!\ldots n_k! }  = \\
&& \sum_{ R \vdash (n_1+ \ldots +n_k)} g( r_1,\ldots,r_k , R ) {d_R \over d_{r_1}\ldots d_{r_k}}
\left ( - \sum_{ \subybox \in R } c_{ \subybox} ( R ) + \sum_{i=1}^k \sum_{ \subybox \in r_i } c_{ \subybox} ( r_i )     \right )^2 \nonumber
\eea
This can be equivalently rewritten as
\bea
\label{newIdn}
&& (\sum_{i=1}^{k-1} \; \sum_{j=i+1}^{k} n_{i} n_{j}) {(n_1 + \ldots + n_k)!   \over n_1!\ldots n_k! } d_{r_1}\ldots d_{r_k} = \\
&& \sum_{ R \vdash (n_1+ \ldots +n_k)} g( r_1,\ldots,r_k , R ) d_R
\left ( - \sum_{ \subybox \in R } c_{ \subybox} ( R ) + \sum_{i=1}^k \sum_{ \subybox \in r_i } c_{ \subybox} ( r_i )     \right )^2 \nonumber
\eea
Wee see that this is a generalization of (\ref{newId}).

\section{ Permutation centralizer algebras, Composite gauge invariant operators and AdS/CFT}\label{sec:Discussion}

The identities above have been derived by calculating the trace in tensor
spaces of some elements in  the group algebra of $ S_{ n_1 + n_2 + \cdots + n_k }$, which are invariant
under conjugation of by permutations in $ S_{ n_1 } \times S_{ n_2 } \times \cdots S_{ n_k}$.
Let us specialise to the case $ k =2$.  The  subspace of $ \mC ( S_{ m + n } )$
which is invariant under conjugation by $ S_m \times S_n$ forms an algebra which has been
studied in detail in \cite{PCA,Kimura}. The motivation came from the role these played in the construction of bases of gauge invariant operators which diagonalise an inner product coming from free quantum field theory \cite{KR1,BHR1,BCD1,BCD2,BHR2}. Key insights into the construction of these bases came from the physics of strings attached to  branes in the context of the AdS/CFT correspondence \cite{CJR,BHLN02,BBFH05}.

Consider quantum fields $ X , Y $ which are $ N \times N $ matrices
 transforming in the adjoint of a $ U(N)$ gauge symmetry.
 \bea
&&  X \rightarrow U X U^{ \dagger } \cr
&&  Y \rightarrow U Y U^{ \dagger}
 \eea
For large $N$,  the space of gauge invariant operators is in 1-1 correspondence with the
 elements of $ \cA ( m , n )$. One way to count the dimension of this space is to  count the traces, which amounts to counting cyclic words built from two letters.
  As  explained in the references above (and reviewed in \cite{review})
  the dimension of the space of gauge invariant operators is also given in terms of
  Littlewood-Richardson coefficients $ g ( R_1 , R_2 , R_3 ) $
  which are multiplicities for the $U(N)$ representation associated with Young diagram $R_3$ (having
  $m+n$ boxes) to appear in the tensor product of $ R_1 \otimes R_2$, where $R_1$ and $R_2$ have
   $m$ and $n$ boxes.
   \bea
   Dim ( \cA ( m , n ) ) = \sum_{ \substack { R_1 \vdash m \\ R_2 \vdash n }} \sum_{ R_3 \vdash m +n }
    ( g ( R_1 , R_2 , R_3 ) )^2
   \eea
The finite $N $ counting is given simply by restricting $R_3$ to have no more than $N$ rows.
This follows by application of Schur-Weyl duality.
The reason these permutation equivalences arise in constructing gauge invariants
is that if we consider a general operator
\bea
X^{ i_1 }_{ j_1 }  \cdots X^{ i_n }_{ j_n}  Y^{ i_{ n+1} }_{ j_{ n+1} }
 \cdots Y^{ i_{ n+m} }_{ j_{ n+m} }
\eea
the upper indices transform in $ V^{ \otimes (m+n) } $ of $U(N)$.
The lower indices transform as $ \bar V^{ \otimes (m+n)} $. The invariants
of $U(N)$ are obtained by contracting with Kronecker $\delta$'s.
As a result, we can construct a gauge invariant for every permutation $ \sigma \in S_{ m+n}$.
\bea
\cO_{ \sigma } ( X , Y ) = X^{ i_1 }_{ i_{ \sigma (1) }  }  \cdots X^{ i_n }_{ i_{ \sigma(n)}} }
 Y^{ i_{ n+1} }_{ i_{ \sigma ( n+1) } }
 \cdots Y^{ i_{ n+m} }_{ i_{ \sigma ( n+m) }
\eea
The bosonic symmetry leads to an equivalence
\bea
\cO_{ \sigma } ( X , Y  ) = \cO_{ \gamma \sigma \gamma^{-1}  } ( X , Y )
\eea
for all $ \gamma \in S_m \times S_n$.
Fourier transformation on $ \cA ( m , n )$ using representation theory of symmetric groups leads to
a Young diagram  basis $ Q^{ R_3 }_{ R_1 , R_2 ; \nu_1 , \nu_2  } $, with
$ 1 \le \nu_1 , \nu_2  \le g ( R_1 , R_2 , R_3 ) $. Representation theoretic formulae for
the Fourier coefficients giving the transformation from trace basis to the Young diagram basis are
given in the papers above. Structural questions about $ \cA ( m ,n ) $, notably regarding minimal sets of generators for maximal commuting sub-algebras, are related to the question of how many charges (generalized Casimirs) are needed to specify a state in the 2-matrix system \cite{EHS,PCA}.  This can be considered to be a measure of complexity of this state space.

$ \cA ( m , n )$ is an example of a permutation centralizer algebra.
$ \cA ( n_1 , \cdots , n_k )$ is analogously defined, and is relevant to gauge
 invariant operators made from $k$ flavours of matrix quantum fields.
 The elements we have used to get the identities above are in fact special
  elements which are central in $ \cA ( n_1, \cdots , n_k )$.
  The central subspace is spanned by products of elements from
   the centre of $ \mC ( S_{ \sum_{ i } n_i } ) $ with elements
    from the centre of $ \prod_i \mC ( S_{ n_i } )  $. (these properties of the centre of PCAs are explained in \cite{PCA} and the special role of the centre in terms of the complexity of
    correlator computations is discussed).      The traces of central elements can thus be obtained using character formulae for      symmetric groups \cite{Lasalle}.

In fact any central element of $ \cA ( n_1 , \cdots , n_k )$ will lead to an identity of
the kind we discussed in the earlier sections.

We may make a few remarks about the analogies which are emerging between quantum information
processing and gauge invariant composite quantum fields through the shared feature of permutation centralizer algebras and Schur-Weyl duality. In QIP, multiple uses of a unitary operation
occur in the computational tasks like % of reconstructing a unitary $U$
oracle based algorithms, estimation problems or arbitrary protocols, which should perform equally well for all states or channels
%\footnote{ For a broad discussion of Schur-Weyl duality in quantum information tasks, see \cite{Hayashi}}.
\footnote{ For a broad discussion of group techniques in quantum information tasks, see \cite{Hayashi}. For use of symmetries in multi step quantum protocols see \cite{comb1,gw2007,comblong,bisioActa}}.
In QFT, these multiple uses of a unitary $U$ arise through the action on a polynomial composite quantum field.

   We thus have a first simple interesting analogy, somewhat simplified from the above set-up,
    which seems to hold promise of    wider implications : \\

{ \it A unitary quantum channel  is analogous to a unitary gauge transformation of an  elementary quantum field } \\

In the QIP problem, multiple uses of channels occur within  multi step quantum protocols  (i.e. within networks of quantum channels). In the composite operator problem of
 QFT, multiple uses occur in different copies of the elementary quantum field
 occuring within a composite. We thus have a second simplified analogy  to think about. \\

 { \it A  multi step quantum protocol is analogous to a composite local operator } \\

The simplicity of these analogies seems to suggest there should be wider applications.
For example, for the multi-partite generalization in Section \ref{multipartite} we may ask, is there
an appropriate optimization task in quantum information theory involving multiple quantum devices  interacting with each other in some way, which employs the multi-partitite  identities (\ref{newIdn}) - generalizing the use of (\ref{desiredId}) in perfect probabilistic storage  and  retrieval \cite{BSZ}?

As noted earlier in this section, structural questions about PCAs have been used to characterize  the complexity of quantum states in  multi-matrix systems, which have a Young diagram basis as well as a trace basis.  The Schur-Weyl  duality transformation from tensor product basis to the Young diagram basis
for $V_N^{ \otimes n } $ has been studied from a quantum information perspective \cite{Harrow}. The question of efficient quantum circuits  having polynomial number of gates has been addressed.  Similar questions can be studied  for the
transformation from trace basis to Young diagram basis for multi-matrix systems. The definition of %efficiency
 complexity  of quantum circuits requires a choice of a basic gate set. A reasonable choice  in the context of AdS/CFT  %or QFT?????
would be to consider the quantum  dilatation operator at one loop and higher loops (see \cite{Beisert03} for the 1-loop dilatation operator and \cite{GiGrav,DoubCos} for applications to brane physics of the action of the one-loop dilatation operator on the Young diagram basis).
 A challenge would be to identify an AdS/CFT dual for such a notion of circuit complexity involving the quantum dilatation operator in the 2-matrix system.

As we have seen, permutation centralizer algebras, with their traces illuminating aspects of perfect probabilistic storing/retrieving and their structure constants having information about correlators of relevance to AdS/CFT,   provide an  intriguing mathematical connection between quantum information and AdS/CFT.  An interesting question is whether there is a physical interpretation of this mathematical connection between QIP and AdS/CFT. In this connection, it is worth noting that studies of quantum state spaces in AdS/CFT from information theoretic perspectives have been undertaken
\cite{IILoss,Simon2018,BM1702}, primarily in the context of state spaces associated with
invariants of a single matrix and  the related free fermion system.
More broadly on  this  theme  the work of
\cite{RT06}  has motivated a rich exploration of connections
 between AdS/CFT and quantum information.
For example it has led to the idea of space-time emerging from entanglement \cite{MVR}
with implications for AdS/CFT holography \cite{FGHMV} and black hole physics \cite{MS}.

{\vskip 0.5cm}
\noindent
\begin{centerline}
{\bf Acknowledgements}
\end{centerline}

SR is supported by the STFC consolidated grant ST/L000415/1 ``String Theory, Gauge Theory \& Duality''
and  a Visiting Professorship at the University of the Witwatersrand, funded by a Simons Foundation
grant to the Mandelstam Institute for Theoretical Physics. SR thanks the Galileo Galilei Institute for Theoretical Physics for  hospitality and the INFN for partial support during the completion of this work. He also thanks the organizers of the  workshop on Matrix Models for non-commutative geometry and string theory in Vienna and the KEK theory group for hospitality  during the completion of this project.  We are grateful for useful discussions to David Berenstein, Robert de Mello Koch, Costis Papageorgakis, Rodolfo Russo, Masaki Shigemori. MS acknowledges the support by the QuantERA project HIPHOP (project ID 731473), projects QETWORK (APVV-14-0878), MAXAP (VEGA 2/0173/17), GRUPIK (MUNI/G/1211/2017) and GA\v CR No. GA16-22211S. MS is grateful to A. Bisio and M. Ziman for fruitful discussions and collaborative work, which led to formulation of the identity (\ref{EQ1}), which is re-derived and generalized in this manuscript. We are grateful to the organizers of the Quantum Physics and Logic (QPL2017) conference
where this interdisciplinary collaboration was initiated.

%% MS acknowledgements still not completely ready

\vskip2cm

\begin{appendix}

\section{some facts about $U(N) $, $S_n$ and the tensor product $V_N^{ \otimes n } $ }\label{sec:App}

This section is brief review of some key facts about the representation theory of symmetric groups, Unitary groups and their relations following from Schur-Weyl duality.
More details are in mathematical physics references such as \cite{Hamermesh} or mathematics texts such as
\cite{FulHar}.
We will start with a useful piece of notation. We will use
 $ r \vdash n $ to denote a partition $r$ of $n$.  Partitions of $n$ correspond to Young diagrams with $n$ boxes, which have row lengths $r_1 \ge r_2 \ge \cdots $, with $ n = r_1 + r_2 + \cdots $.
Young diagrams with $n$ boxes correspond to irreducible representations of $S_n$.
Letting $V_N$ be  the fundamental representation of $U(N)$, the tensor product $V_N^{ \otimes n }$
is a representation of the diagonal $U(N)$ acting as
\bea
U \otimes U \otimes \cdots \otimes U
\eea
as well as the symmetric  group of  all permutations of $n$ objects ($S_n$). These two actions commute with each other, which leads to Schur-Weyl duality
\bea
 \label{eq:schurweyl1}
  V_N^{ \otimes n } = \bigoplus_{ r \vdash n  } V_r^{ U(N)} \otimes V_r^{ S_n }
\eea
This gives the decomposition of $ V_N^{ \otimes n } $ into irreducible reps of
$U(N) \times S_n$ as a direct sum labelled by Young diagrams.

  A useful  formula for the dimension of unitary group  $U(N)$ irreps in terms of characters of $S_n$ is
  \bea
  { 1 \over n! } \sum_{ \sigma \in S_n } \chi_{ r } ( \sigma ) N^{ C_{ \sigma } }  = \Dim_N r \; ,
  \eea
  where $ C_{ \sigma } $ is the number of cycles in the permutation $ \sigma $.
 This follows from Schur-Weyl duality  (\ref{eq:schurweyl1}).
  To project to a fixed Young diagram, we can use a projector  element in the
   group algebra
   \bea
   P_r = { d_r \over n! } \sum_{ \sigma \in S_n } \chi_r ( \sigma ) \sigma
   \eea
   If we apply this to the states in $ V_{N}^{ \otimes n } $ and take a trace, we need to calculate
   \bea
&& tr_{  V_N^{ \otimes  n } } ( \sigma ) =
\langle e_{ i_1} \otimes \cdots \otimes e_{ i_n } | \sigma |   e_{ i_1} \otimes \cdots \otimes e_{ i_n } \rangle \cr
&&  = \langle e_{ i_1} \otimes \cdots \otimes e_{ i_n }  |   e_{ i_{ \sigma( 1) } } \otimes \cdots \otimes e_{ i_{ \sigma(n)}  } \rangle \cr
&& = N^{ C_{ \sigma } }
   \eea
    ( usual summation convention, so the $i$ indices are summed from $1$ to $N$).
   To understand the last line, it is instructive to do some examples at $n=2$.
   If $ \sigma = (1) (2) $, the trace is
   \bea
   \delta_{ i_1 , i_1} \delta_{ i_2 , i_2 }  = N^2
   \eea
   If $ \sigma = ( 12) $, the trace is
   \bea
   \delta_{ i_1 , i_{ \sigma (1)} }   \delta_{ i_2 , i_{ \sigma (2)} }
    = \delta_{ i_1 , i_2} \delta_{ i_2 , i_1}  = N
   \eea

We need to understand  some  multiplications in the group algebra
 of $S_{ n+1}$. The group algebra consists of formal sums of group elements
  with complex coefficients.
  What happens when a generic group element $ \sigma $  in the $S_n$ subgroup
   is multiplied with $ ( a , n+1)$ for $ a \in \{ 1, \cdots , n \}$ ?
    Example at $n =3$, with $ \sigma = ( 1,2,3) $
    \bea
    ( 1,2,3) ( 2, 4 ) = ( 1,4,2,3)
    \eea
The number of cycles in $ \sigma . ( a , n +1) $ is the same as the
 number of cycles in $ \sigma $. As a result, if
 \bea
&& \tau =  \sigma . \sum_{ a =1}^{ n} ( a , n+1)  \cr
&& C_{ \tau } = C_{ \sigma }
 \eea
This implies that
\bea\label{trsumprod}
&& tr_{ V_N^{\otimes n+1} } ( {P_r \over d_r }  \sum_{ a =1}^{ n } ( a , n+1) )
= { 1 \over n! }\sum_{ a =1}^{ n } \sum_{ \sigma \in S_n }  \chi_r ( \sigma ) tr_{ V_N^{\otimes n+1} } \left ( \sigma .  ( a , n+1) \right )  \cr
&& =  \sum_{ a =1}^{ n } \Dim_N r = n \Dim_N r
\eea

Central elements ( such as $T_2^{ (S_n)} $) multiplying a projector
give normalized characters times the projector.
\bea\label{centralchar}
T_2^ {(S_n)} P_r =  { \chi_r ( T_2^{ S_n} ) \over d_r } P_r
\eea
To see this, note that both LHS and RHS are central elements in the group algebra of $ S_n$, as a result
they are determined by their irreducible characters, and we can easily verify that the two sides have the same irreducible characters.
The normalized character is known \cite{Lasalle} to be the sum
of contents
\bea\label{contents}
 { \chi_r ( T_2^{ S_n} ) \over d_r } = \sum_{ \ybox \in r } ( - c_{ \subybox} ( r )  )
\eea

An implication of Schur-Weyl duality is that the  Littlewood-Richardson coefficients  $ g( r_1 , r_2 ; R )$ which give the multiplicities of $U(N)$ tensor product decompositions
\bea
V_{r_1}^{ U} \otimes V_{ r_2}^U = \oplus_{ R } g ( r_1 , r_2 , R ) V_R^U
\eea
also have an interpretation purely in terms of symmetric groups. They are the reduction multiplicities
for the decomposition of the irrep $V_R^{ S_n } $ in terms of the subgroup $S_{n_1}\times S_{ n_2} $.
We may express this as
\bea
V_R^{ S_n} = \bigoplus_{ r_1 , r_2 } V^{ S_{n_1}}_{r_1} \otimes V^{S_{n_2} }_{ r_2} \otimes V_{ r_1 , r_2 }^R
\eea
where $ V_{ r_1 , r_2 }^R  $ is the multiplicity space, of dimension $ g( r_1, r_2 , R ) $.
Considering the trace in $ V_R^{ S_n}$
\bea
tr ( P_{ r_1 } \circ P_{ r_2} )
\eea
we arrive at
  \bea\label{LRchar}
  g ( r_1  , r_2 , R ) = { 1 \over n_1 ! n_2 ! } \sum_{ \sigma_1 \in S_{ n_1} } \sum_{ \sigma_2 \in S_{ n_2} } \chi_{ r_1} ( \sigma_1 ) \chi_{r_2} ( \sigma_2 ) \chi_R ( \sigma_1 \circ \sigma_2 )
  \eea
  More generally
  \bea
  V_{r_1}^{ U} \otimes V_{ r_2}^U \otimes \cdots  \otimes V_{ r_k}^U
  = \oplus_{ R } ~~  g ( r_1 , r_2 , \cdots r_k ;  R ) V_R^U
  \eea
  and
  \bea\label{keyId}
   g ( r_1 , r_2 , \cdots r_k ;  R ) = \sum_{ \sigma_1 \in S_{ n_1} } \cdots  \sum_{ \sigma_k \in S_{ n_k } }
   \left (  \prod_{ i=1}^{ k } { \chi_{ r_i} ( \sigma_i  ) \over n_i! } \right )
    \chi_R ( \sigma_1 \circ \sigma_2 \circ \cdots \circ \sigma_k )
  \eea

Let $ W = V_N^{ \otimes n  } = V_{ N}^{ \otimes ( n_1 + n_2  + \cdots + n_k ) } $.
Consider the trace
\bea
tr_{ W} P_R \left (   { P_{ r_1} \over d_{ r_1} }  \otimes  { P_{ r_2} \over d_{ r_2} } \otimes
\cdots  \otimes { P_{ r_k} \over d_{ r_k} }  \right )
\eea
where $ R $ is a Young diagram with $n $ boxes, $r_i$ are Young diagrams with $n_i$ boxes, $P_R$ and $P_{r_i} $ are the corresponding  projectors.%boxes.
Using Schur-Weyl duality
\bea
W = V_{N}^{ \otimes n } = \bigoplus_{ R \vdash n } V_R^{ U }  \otimes V_R^{ S_n }
\eea
When the projector $P_R$ acts on $W$, we project to a single factor $V_R^U  \otimes V_R^{ S_n } $.
We can decompose $ V_R^{ S_n } $ in terms of $S_{ n_1} \times S_{ n_2} \times \cdots \times S_{ n_k} $. The multiplicities are the Littlewood-Richardson coefficients.
\bea
V_{ R}^{ U}  \otimes V_R^{ S_n}
= V_R^U \otimes \bigoplus_{ r_1 , r_2 , \cdots , r_k } g( r_1 , r_2 , \cdots , r_k ; R)   \left ( V_{ r_1}^{ S_{ n_1} } \otimes V_{ r_2}^{ S_{ n_2} } \otimes \cdots \otimes V_{ r_k}^{ S_{n_k} }  \right )
\eea
It follows that
\bea
tr_{ W} P_R \left (   { P_{ r_1} \over d_{ r_1} }  \otimes  { P_{ r_2} \over d_{ r_2} } \otimes
\cdots  \otimes { P_{ r_k} \over d_{ r_k} }  \right )
= ( Dim_N R ) g ( r_1 , r_2 , \cdots , r_k ; R )
\eea
This is an important identity we use in the paper.
To make the above proof more explicit, we can expand the projectors in terms of characters.
\bea
&& tr_{ W} P_R \left (   { P_{ r_1} \over d_{ r_1} }  \otimes  { P_{ r_2} \over d_{ r_2} } \otimes
\cdots  \otimes { P_{ r_k} \over d_{ r_k} }  \right )  \cr
&& = \sum_{ \sigma_1 \in S_{ n_1} } \cdots \sum_{ \sigma_k \in S_{ n_k} } \sum_{ \sigma \in S_n }
{ d_R \chi_R ( \sigma ) \over n! }  \left (  \prod_{ i =1}^{ k } {  \chi_{ r_i} ( \sigma_i ) \over n_i! }   \right )
tr_W \left ( \sigma ( \sigma_1 \circ \sigma_2 \circ \cdots \circ \sigma_k ) \right )  \cr
&&
\eea
For any permutation $ \tau \in S_n$, Schur-Weyl duality implies that
\bea
tr_W ( \tau ) = \sum_{ S \vdash n } \chi_S ( \tau )  ~ Dim_N S
\eea
Hence
\bea
&& tr_{ W} P_R \left (   { P_{ r_1} \over d_{ r_1} }  \otimes  { P_{ r_2} \over d_{ r_2} } \otimes
\cdots  \otimes { P_{ r_k} \over d_{ r_k} }  \right )   \cr
&& = \sum_{ \sigma_1 \in S_{ n_1} } \cdots \sum_{ \sigma_k \in S_{ n_k} } \sum_{ \sigma \in S_n }
{ d_R \chi_R ( \sigma ) \over n! }  \left (  \prod_{ i =1}^{ k } {  \chi_{ r_i} ( \sigma_i ) \over n_i! }   \right ) \sum_{ S \vdash n } \chi_S ( \sigma (  \sigma_1 \circ \sigma_2 \circ \cdots \circ \sigma_k ) )
Dim_N S \cr
&&
\eea
Using the character orthogonality relation
\bea
  { 1 \over n! } \sum_{ \sigma } \chi_{ S } ( \sigma ) \chi_R ( \sigma \rho )  = \delta_{ R S } { \chi_R ( \rho ) \over d_R }
\eea
which holds for any $ \rho \in S_n $, we have
\bea\label{KeyId}
&& tr_{ W} P_R \left (   { P_{ r_1} \over d_{ r_1} }  \otimes  { P_{ r_2} \over d_{ r_2} } \otimes
\cdots  \otimes { P_{ r_k} \over d_{ r_k} }  \right ) \cr
&& =
  \sum_{ \sigma_1 \in S_{ n_1} } \cdots \sum_{ \sigma_k \in S_{ n_k} } \sum_{ \sigma \in S_n }
   \left (  \prod_{ i =1}^{ k } {  \chi_{ r_i} ( \sigma_i ) \over n_i! }   \right )
    \chi_R (  \sigma_1 \circ \sigma_2 \circ \cdots \circ \sigma_k ) Dim_N R \cr
&& = g ( r_1 , r_2, \cdots , r_k ; R )  Dim_N R
\eea

\end{appendix}

\end{document}